\documentclass[12pt]{article}
\newcommand{\beq}{\begin{equation}}
\newcommand{\eeq}{\end{equation}}
\newcommand{\beqa}{\begin{eqnarray}}
\newcommand{\eeqa}{\end{eqnarray}}

\def\rad{\mathcal{T}}
\def\gsim{\mathrel{\rlap{\lower4pt\hbox{\hskip1pt$\sim$}}
    \raise1pt\hbox{$>$}}}                

\long\def\symbolfootnote[#1]#2{\begingroup%
\def\thefootnote{\fnsymbol{footnote}}\footnote[#1]{#2}\endgroup}

\newif\ifdohyperref
\dohyperreffalse

\dohyperreftrue
\ifdohyperref
  \usepackage[hyperindex]{hyperref}
 \usepackage[comma,square]{natbib}
  \newcommand\email[1]{\href{mailto:#1}{#1}}
  \newcommand\eprint[1]{\href{http://arXiv.org/abs/#1}{[arXiv:#1]}}
\else
 \usepackage[sort&compress,comma,square]{natbib}
  \newcommand\email[1]{#1}
  \newcommand\eprint[1]{[arXiv:#1]}
\fi

\begin{document}

\thispagestyle{empty}
\vspace{1mm}
\begin{center}
\vspace*{1cm}
{\Large \bf Supersymmetry breaking and the radion in AdS$_4$ brane-worlds}\\
\vspace*{0.8cm}
{\large Andrey Katz$^a$, 
Yael Shadmi$^a$, and
Yuri Shirman$^b$}\symbolfootnote[0]{\email{andrey@physics.technion.ac.il},
\email{yshadmi@physics.technion.ac.il},     
\email{shirman@lanl.gov}}\\
\vspace{0.8cm}
$^a$ {\em Physics Department, Technion, Haifa 32000, Israel}\\
$^b$ {\em T-8, MS B285, LANL, Los Alamos, NM 87545}\\
\vskip 0.2in
\vspace{0.7cm}
ABSTRACT
\end{center}
We compute the one-loop correction to the radion potential
in the Randall-Sundrum model with detuned brane tensions,
with supersymmetry broken by boundary conditions.
We concentrate on the small warping limit, where
the one-loop correction is significant. 
With pure supergravity, the correction is negative,
but with bulk hypermultiplets, the correction can be positive,
so that the 4d curvature can be lowered, with the radion stable.
We use both the KK theory, and the 4d radion effective theory
for this study.
\setcounter{page}{0} \setcounter{footnote}{1}
\newpage


\section{Introduction}
It is well known by now that a 3-brane in AdS$_5$
can localize gravity~\cite{Karch:2000ct},
even when the brane tension is not tuned to the bulk cosmological constant
as in the Randall-Sundrum (RS) model~\cite{Randall:1999ee,Randall:1999vf}.
The 4d theory can be either dS$_4$ or AdS$_4$
near the brane, with the 4d curvature proportional to
the amount of detuning.
In the latter case, 
if a second brane is added to the theory, 
the jump conditions for this brane constrain its position,
so that the brane distance is determined
by the bulk cosmological constant and two brane 
tensions~\cite{Kaloper:1999sm}, and, unlike in the RS model,
the radion is stabilized.

Another qualitative difference between detuned and tuned
brane systems has to do with supersymmetry breaking. 
Supersymmetric extensions of detuned brane models
involve gravitino brane mass-terms, whose magnitudes are 
proportional to the amount of detuning~\cite{Bagger:2002rw}.
When the phases of these two brane-terms differ,
supersymmetry is broken~\cite{Bagger:2003vc}.
This is clearly impossible in the tuned limit, 
where the brane terms vanish and their phases
are ill-defined.
Thus, the detuned theory, even with just pure gravity in the bulk,
allows for spontaneous supersymmetry breaking,
with the radion automatically stabilized.
But these desirable features come at a price:
the 4d theory is AdS$_4$, and so the 4d cosmological constant has the
wrong sign for phenomenological applications.
For model building purposes, it is therefore important to find
modifications of the model that can lift the theory to Minkowski space
without upsetting the radion stabilization.
This has to be
achieved while at the same time stabilizing the scalar 
partner of the radion, which, as we will explain, 
is a modulus of the classical theory.

In this paper, we will study the effects of supersymmetry-breaking
quantum corrections on the radion potential.
These effects are finite and calculable because the supersymmetry-breaking
mechanism described above is non-local.
While gravity gives a negative correction to the potential,
hypermultiplets contribute with the opposite sign, 
so that the net contribution can offset, at least
partially, the negative cosmological constant.

We focus here on models in which the
5d curvature, $k$, is much smaller than the inverse 
brane distance $1/R$.
This complements  the analysis of~\cite{us}, which mostly considered
the large warping case.
For small warping, the loop correction to the potential
can be important. 
This correction  depends on two
dimensionless parameters: the warp factor $kR$, and 
the 4d curvature in units of $k$, $1/(kL)$, which is proportional
to the amount of detuning.
Keeping $kL$ fixed and decreasing the warping, the loop correction should
reproduce the non-zero Casimir energy of flat orbifold models
with broken supersymmetry.
Thus, while the classical potential decreases for small warping, 
the loop correction does not.
Furthermore, because it is a non-local effect, the supersymmetry
breaking correction to the potential involves the warp factor 
$exp(-k \pi R)$, and is therefore suppressed for large warping.
Note that since our framework is supersymmetric, there is no need for large
warping in order to generate the weak-Planck scale hierarchy.
Rather, our analysis is relevant for the problem of radius stabilization
in models with an almost-flat extra dimension.

Our approach is complimentary to the approach
of radion stabilization models based on the Casimir 
energy~\cite{Garriga:2000jb,Goldberger:2000dv,Ponton:2001hq,
Garriga:2002vf}. 
While these typically start with zero radion potential
at the classical level and use the Casimir energy in order
to stabilize the radion, here we start with a small 
classical potential with the radion stabilized,
and consider the Casimir energy as a correction to 
the cosmological constant.
In the supersymmetric 4d theory, the radion is accompanied
by the fifth component of the graviphoton,
which is a modulus of the classical theory,
even in the detuned case.
As we will see, this field, which we will call $b$,
is stabilized by the Casimir energy.
In fact, because the graviphoton gauges a $U(1)_R$
symmetry of the 5d theory, under which the gravitino
is charged, any phase difference of the gravitino
brane terms can be compensated by a non-zero $b$,
so the $b$ vacuum expectation value (VEV) breaks supersymmetry~\cite{brwilson}.
At one-loop, $b$ is stabilized either at the origin,
corresponding to unbroken supersymmetry, or for
maximal supersymmetry breaking.

There have been many studies of supersymmetry-breaking
in brane worlds with AdS$_5$ bulk in the past few years
(see for example~\cite{Luty:2000ec,Luty:2002hj,BB,Maru:2005qx}). Some of these
start with the tuned RS case but invoke additional sources of bulk
and brane energies (from, say, a supersymmetry breaking sector),
or brane gravitino mass terms, and
rely on a {\it supersymmetric} 4d effective description.
However, the unbroken 4d supersymmetry is determined by the
bulk cosmological constant and brane tensions. Boundary terms
can break this supersymmetry spontaneously if they are in certain
ranges, but otherwise, as in the dS$_4$ case, 
break it explicitly\footnote{As is the case for ref.~\cite{Maru:2005qx}.}.
Our starting point in this paper is the supersymmetric 5d theory
of~\cite{Bagger:2002rw,Bagger:2003vc}, for which supersymmetry breaking
is well understood. At low energies, this theory gives rise to a
supersymmetric effective action for the radion supermultiplet, with a
known superpotential and K\"ahler potential~\cite{Bagger:2003dy}.
We will use the two descriptions to calculate the potential
at one-loop.
In both cases, we assume  small detunings, so that the 4d curvature, $1/L$,
is small compared to the 4d Planck mass, $M_4$, and work to leading order in
$1/L$.

Starting from the 5d theory in section~\ref{kksection}, 
we compute the shifts of the gravitino KK masses due to supersymmetry 
breaking.
Because the contribution of a full KK supermultiplet to the Casimir energy
vanishes for unbroken supersymmetry, these shifts allow us to derive
the correction to the potential.
Furthermore, these mass shifts are proportional to $1/L$,
because, as explained above, supersymmetry can only be broken in
the detuned case.
To leading order in the 4d curvature we can therefore use
flat-space propagators in the calculation, avoiding the complications
of a full curved space calculation~\cite{Saharian:2002bw,Norman:2004qt}.

In section~\ref{4dsection} we turn to the 4d radion effective theory.
The superpotential of this theory is proportional 
to $1/L$~\cite{Bagger:2003dy}
and is not corrected at one-loop.
Therefore, to obtain the leading order contribution to the potential,
only the tuned (RS) K\"ahler potential is needed.
The tree-level K\"ahler potential was obtained
in~\cite{Luty:2000ec,Bagger:2003dy}
and one-loop corrections were calculated
in~\cite{Buchbinder:2003qu,Rattazzi:2003rj,Gregoire:2004nn,Falkowski:2005fm},
so we can just use known 
results.
The 4d radion theory also allows us
to calculate the contribution of hypermultiplets to the potential.
It would be interesting to incorporate hypermultiplets in the 5d
supersymmetric theory, but we leave this for future work.
Still, there is no reason to expect that the mere addition of
hypermultiplets to the theory would
break supersymmetry, and so it is safe to study them using the
supersymmetric 4d effective theory.
Again, only the RS K\"ahler potential is needed,
and the 1-loop hypermultiplet contribution to this K\"ahler potential 
was discussed in~\cite{Falkowski:2005fm}.

\section{Radion potential: KK calculation}\label{kksection}
We consider pure supergravity on a 5d orbifold,
with AdS$_5$ bulk, and brane tensions detuned from the RS limit so 
that the 4d
slices are AdS$_4$~\cite{Bagger:2002rw}.
(We summarize the elements of this theory in Appendix~\ref{setup} 
for convenience.)
For the 5d action to be supersymmetric, 
gravitino brane mass terms are required.
Labeling the branes by 0 and $\pi$ according to their position
in the 5th dimension, the brane mass terms, $\alpha_i$, are given by,
\beq\label{tensions}
|\alpha_0|^2 = \frac{T-T_0}{T+T_0} = \frac1{4k^2 L^2}\ ,\ \ \ 
{\rm and}\ \  \ \  
|\alpha_\pi|=|\alpha_0|e^{k\pi R} \ ,
\eeq
where $T_0$ is the tension of the brane at zero,
$L$ is the 4d radius of curvature,
and $T\equiv 6 M_5^3 k$, with $M_5$ 
the 5d Planck scale and $k$ the 5d curvature.
The distance between the branes, $R$, is determined by the brane tensions
and $T$ [see eqn.~(\ref{radius})].
Note that supersymmetry gives an upper bound on the magnitudes
of the two brane tensions.

When eqn.~(\ref{tensions}) is satisfied, the action preserves
$\mathcal{N}=1$ supersymmetry locally at any point along the fifth
dimension. Globally however, $\mathcal{N}=1$ supersymmetry is only preserved
when solutions to the Killing spinor equations exist. 
This, in turn, requires that the
phases of $\alpha_0$ and $\alpha_\pi$ are equal. When these phases differ,
supersymmetry is spontaneously broken. 
Since the gravitino is charged under a gauged $U(1)$
subgroup of the $SU(2)_R$ symmetry of the bulk supergravity,
we can rotate away the phase difference by turning on a non-zero
VEV for the graviphoton field---the $U(1)$ gauge boson---in the 
fifth dimension.
The condition for unbroken supersymmetry then becomes
\beq
|\alpha_\pi|e^{i\phi_\pi}=|\alpha_0|e^{i\phi_0} e^{k\pi (R-i
  \sqrt{6}B_5)} \ ,
\eeq
where $B_5$ is the fifth component of the graviphoton.
In this section we will set $B_5=0$ and work with the 
supersymmetry-breaking
parameter 
\beq
\phi=\phi_0-\phi_\pi \ .
\eeq

For unbroken supersymmetry, or $\phi=0$, the contribution of each
KK supermultiplet to the potential vanishes.
A supersymmetry-breaking nonzero $\phi$ only affects the gravitini masses.
Therefore, 
we can calculate the correction to the potential
by considering only the gravitini KK tower,
\beq\label{fermv}
\Delta V(\phi)=
\Delta V_{{\mbox{\tiny{bosons}}}}(\phi)
+  \Delta V_{{\mbox{\tiny{fermions}}}}(\phi)
 =
\Delta V_{{\mbox{\tiny{fermions}}}}(\phi) -
V_{{\mbox{\tiny{fermions}}}}(\phi=0) \ .
\eeq
In order to calculate this correction, we need the KK gravitini
masses, both in the supersymmetric case and with broken 
supersymmetry.
In the tuned RS case, there is a pair of degenerate spin 3/2 states
at each KK level.  
The degeneracy is lifted when the brane tensions are detuned.
In Appendix~\ref{shifts}, we derive the gravitini mass shifts
for small supersymmetry breaking, $\phi\ll 1$.
To leading order in $1/L$, the two masses at each KK level are split
by opposite amounts, and there is no correction to
the potential.
Expanding the masses to the next order in $1/L$ we find 
(see Appendix~\ref{shifts})
\beq\label{masses}
{{m^{(n)}}^{\pm}\over k}= c^{(n)}_0 \pm \frac{1}{kL} \left[ c^{(n)}_1
+ c^{(n)}_{1,SB} \phi^2\right]
+ \frac{1}{(kL)^2} \left[ c^{(n)}_2  + c^{(n)}_{2,SB} \phi^2\right]
+ {\cal O}(\phi^4)\ .
\eeq
The dimensionless coefficients $c$, which depend on $k$ and $R$,
are given in Appendix~\ref{shifts}. 
In particular, $c^{(n)}_0$ is the mass of the $n$-th KK mode
in the tuned RS model in units of $k$.

As expected, the mass shifts due to the supersymmetry breaking
are proportional not just to $\phi^2$, but also
to  the 4d curvature $1/(kL)$.
This reflects the fact that supersymmetry cannot be broken
in the tuned case.
We can therefore use flat-space propagators
for the calculation.
The curved-space propagators are the flat space ones
plus ${\cal O}(1/L^2)$ corrections.
To leading order, if we keep these $1/L^2$
corrections in the propagators, we should keep just the zeroth order
KK masses, and these give a zero result in~(\ref{fermv}).

We can now consider the contribution of the $n$-th KK mode
to the Casimir energy.
As argued above, this contribution is of the
form~(\ref{fermv}), namely, the difference between the
fermion contributions with and without supersymmetry breaking.
Substituting the masses~(\ref{masses}) we have
\beqa\label{vacen}
\Delta V=4 \frac{1}{L^2} \, \phi^2\, \times
\ \ \ \ \ \ \ \ \ \ \ \ \ \ \ \ \ \ \ \ \ \ \ \ \ 
\ \ \ \ \ \ \ \ \ \ \ \ \ \ \ \ \ \ \ \ \ \ \ \ \ 
\ \ \ \ \ \ \ \ \ \ \ \ \ \ \ \ \ \ \ \ \ 
\\
\sum_{n} \int \frac{d^4p}{(2\pi)^4}\,
\Biggl[\,
2\,\frac{k^2 (c_0^{(n)})^2 \, c_1^{(n)}\, c_{1,SB}^{(n)}}
{\left[p^2+ (m_0^{(n)})^2\right]^2} 
-\,\frac{c_0^{(n)}\,  c_{2,SB}^{(n)}}{p^2+(m_0^{(n)})^2}
-\,\frac{ c_1^{(n)}\, c_{1,SB}^{(n)}}{p^2+(m_0^{(n)})^2  }\, \Biggr]~,
\nonumber\eeqa
up to $1/L^4$ terms.
Here $m_0^{(n)}= k c_0^{(n)}$.
Note that this correction is linear in the supersymmetry-breaking
mass shifts, and therefore in the supersymmetry-breaking
scale.
This can be seen clearly in equation~(\ref{vacen}).
One power of $1/L$ in this equation comes from the
fact that the correction is proportional to the 4d curvature,
while the remaining  $\phi^2/L$ comes from the supersymmetry
breaking scale.
However, the supersymmetry breaking scale must also involve
the warp factor $exp(-k\pi R)$ because it is a non-local effect.
Therefore, the correction to the potential can only be significant
for small warping. 
Furthermore, for small $k$ we approach flat space, where
the classical potential for the radion vanishes, 
so radiative corrections 
may be comparable to the tree level terms.
We will therefore choose
\beq\label{flatlim}
\frac{1}{L}\ll k \ll \frac{1}{R} \ .
\eeq
In terms of the 5d scales, this corresponds to taking $k\ll M_5$,
with the ratios $T_0/T$, $T_\pi/T$ held fixed, 
so that the radius $R$ is fixed [see eqn.~(\ref{radius})]
and $1/(kL)$ is a small number.

With this choice we can calculate the correction~(\ref{vacen})
analytically.
We then find (see Appendix~\ref{flat} for details)
\beq\label{kkcorr}
\Delta V = - {3 \zeta(3)\over
2^5\pi^2}
{1\over (\pi R)^4}\, {1\over ( k L)^2} \phi^2\ .
\eeq
As expected, the correction is cut-off by the first KK
mass, which is roughly $1/R$, and goes to zero for zero 4d curvature,
for which the supersymmetry-breaking shifts vanish.
Apart from the overall $1/(kL)^2$,
the correction does not involve $k$:
it appears at zeroth order in the warping $kR$.
This is not surprising, because, as mentioned above,
the Casimir energy is non-zero in flat space in the presence
of supersymmetry-breaking. 

Indeed, the result~(\ref{kkcorr}) reproduces the Casimir energy
of flat orbifold models, with supersymmetry broken by brane
superpotentials~\cite{Bagger:2001ep,Quiros:2003gg}, 
in the limit that the brane superpotentials are small. 
In the flat orbifold models of~\cite{Bagger:2001ep},
the brane superpotentials can be arbitrary.
Here they are dictated by supersymmetry,
and proportional to the detuning, $1/(kL)$, which we take to be small
[see eqn.~(\ref{tensions})].
Note that for small warping, the brane terms can only differ by a phase,
$\alpha_\pi\sim e^{i\phi} \alpha_0$.
The comparison to flat orbifold models is in fact non-trivial,
because the physics is quite different.
In particular the KK spectrum~(\ref{masses}) is different from the spectrum of
flat orbifold models. Effectively however, the KK contributions to the 
potential take the same form in both cases.

Thus, for small warping, the one-loop correction to the potential
is important.
We will return to this point in the next section,
where we discuss the behavior of the one-loop improved
potential starting from the 4d radion theory.

We note that, while we could only calculate $\Delta V$
analytically to leading order in $kR$, 
we can compute it numerically for any warping,
using eqns.~(\ref{startc})-(\ref{endc})
for the gravitino mass shifts.

At this point, we can 
deduce
the mass of the graviphoton field $b$ (defined in terms of $B_5$
according to (\ref{graviphoton})).
As mentioned above, a non-zero $\phi$ can be rotated
away by a VEV of $b$, which is a modulus of the
classical theory.
Given also that $b$ is periodic, with period $2/(3k)$~\cite{Bagger:2003dy},
we can plausibly replace $\phi\to \phi-3\pi kb$
in eqn.~(\ref{kkcorr}).  
This guess will be borne out by the analysis of the next section.

\section{Radion potential: 4d radion theory}\label{4dsection}
We can also evaluate the correction to the energy
using the 4d radion theory following the approach of \cite{us}.
This will allow us to obtain the potential for  finite
values of the supersymmetry breaking phase $\phi$
and to derive its dependence on the graviphoton zero-mode.
It will also allow us to include
the effect of bulk matter fields. 

The effective 4d action for the radion
was found in \cite{Bagger:2003dy} by matching to the 5d theory at
tree-level.
The theory has the
 superpotential\footnote{We rescale the 
superpotential and K\"ahler potential of~\cite{Bagger:2003dy}
as 
$K\to K+3M_4^2\ln(1-e^{-2k\pi R})$,
$W\to \frac{W}{(1-e^{-2k\pi R})^{3/2}
}$.},
\beq\label{super}
W = \frac1{1-e^{-2k\pi R}}\,
{M_4^2\over L} \, \left(1- e^{i\phi}\, e^{k\pi R}\,
e^{-3k\pi \rad}\right)\ ,
\eeq
where $\rad$ is the radion superfield, whose scalar component
is $r+ i b$, where $r$ is the radion and $b$ is the graviphoton Wilson
line (see Appendix~\ref{setup} for their definitions in terms of the 
5d theory).

The superpotential~(\ref{super}) is proportional to $1/L$, and
vanishes when the brane tensions are tuned.
Indeed, this superpotential is essentially
the sum of the two brane superpotentials $\alpha_0$ and
$\alpha_\pi$, ``weighted'' by the appropriate warp factor.
In the tuned case, $\alpha_0=\alpha_\pi=0$ and the superpotential
vanishes.

If we are only interested in the tree-level potential to leading
order in the 4d curvature, we can therefore use the RS
K\"ahler potential
\beq\label{kahler}
K = -3 M_4^2 \ln\left(
\frac{1 - e^{-k \pi (\rad+\bar{\rad})}}{1-e^{-2k\pi R}}
\right)\ ,
\eeq
neglecting ${\cal O}(1/(M_4 L)^2)$ terms.

With this K\"ahler potential and superpotential, the order
parameter for supersymmetry breaking is~\cite{Bagger:2003dy}
\beq\label{superb}
D_\rad W \propto \left(1- e^{i(\phi -3k \pi b)}\right)\ ,
\eeq
so that supersymmetry is broken for $3k \pi b\neq\phi$.
However, it is easy to verify that the tree-level potential
is independent of 
the phase in~(\ref{superb}).
This is a very peculiar theory:
Its K\"ahler potential is that of no-scale supergravity,
and similarly its potential is $b$-independent.
But whereas the no-scale model has a constant superpotential,
and supersymmetry broken everywhere, here the superpotential
is $b$-dependent, and supersymmetry remains unbroken for 
$3k \pi b = \phi$.

At the loop level, the superpotential is not corrected.
This is not quite the familiar ``non-renormalization''
of the superpotential---in fact, the superpotential~(\ref{super}), which is
obtained by matching to the 5d theory at the KK scale, involves
$M_4^2$, $R$, $L$, and $k$, all of which are
physical, renormalized quantities.
To obtain the supersymmetric low-energy radion theory
at one-loop, one should again match to the 5d theory,
this time at one-loop.
As mentioned above, the superpotential~(\ref{super})
is essentially the sum of the two brane-superpotentials,
weighted by the warp superfield $exp(-3 k\pi \rad)$.
Since the brane superpotentials are not renormalized,
and the warping is dictated by the symmetry of the
5-dimensional space, 
the superpotential~(\ref{super}) remains unchanged.

To compute the potential at one-loop we therefore just need
the one-loop correction to the K\"ahler potential.
Furthermore, to leading order in $1/L$, 
the K\"ahler potential 
is just the RS K\"ahler potential, 
which depends only on the combination
$\rad+\bar{\rad}$~\footnote{At tree-level, this holds to all orders
in $1/L$. At the loop-level, the K\"ahler potential will
have terms proportional to $1/L^2$. Such terms are suppressed
compared to the terms we discuss here by $1/(M_4 L)^2$.
}.
For $K=K(\rad+\bar{\rad})$ we can derive a simple general formula for
the potential, 
\beqa
\label{eq:potential}
V&=& e^{K/M_4^2}\,\frac{\vert W_0\vert^2}{M_4^2}
\Bigg\{ M_4^2\, K_{\rad\bar{\rad}}^{-1} 
\left[\frac{K_\rad}{M_4^2} +e^{\pi k(R-3r)}\, (3\pi k-\frac{K_\rad}{M_4^2})
\right]^2\nonumber\\
&-&3 \left[1-e^{\pi k(R-3r)}\right]^2\\ 
&-& 4 e^{\pi k(R-3r)} \sin^2{3\pi kb-\phi\over2} 
\left[ \frac{K_\rad}{K_{\rad\bar{\rad}}} (3\pi k-\frac{K_\rad}{M_4^2}) 
+3\right]
\Bigg\}\ ,\nonumber
\eeqa
where $W_0=M_4^2/((1-e^{-2k\pi R})L)$. 
Writing $K=-3M_4^2 \log(\Omega/3M_4^2)$,
the supersymmetry-breaking, or $b$-dependent part of the potential
simplifies further, 
\beq
\Delta V=-  4\cdot 3^4\, \vert W_0\vert^2 M_4^4\, e^{-2\pi k R}\,
\frac{\Omega^\prime +\Omega^{\prime\prime}}{\Omega^2
(\Omega^{\prime\prime}\Omega -(\Omega^\prime)^2)}\,
\sin^2{3\pi kb-\phi\over2}\,,
\eeq
where the primes denote derivatives with respect to 
$k\pi \rad$~\footnote{With a slight abuse of notations
we use double prime to refer to the derivative with respect
to $\rad$ and $\bar\rad$. Since $\Omega=\Omega(\rad+\bar\rad)$,
this is the same as the second derivative with respect to $\rad$.}.
At the tree-level, $\Omega^\prime +\Omega^{\prime\prime}=0$,
so we can set $\Omega=\Omega_{tree}$ everywhere else to obtain, up to
terms of the order $\mathcal{O}(1/(M_4L)^4)$,
\beq
\Delta V=   4\, e^{2\pi kR} 
\frac1{L^2}\,
\left(\Delta\Omega^\prime +\Delta\Omega^{\prime\prime}\right)
\, \sin^2{3\pi kb-\phi\over2}\ ,
\eeq
where $\Delta\Omega$ is the one-loop correction to $\Omega$.

The potential is therefore extremized 
for either zero or maximal supersymmetry breaking.
Which one of these is the minimum/maximum
depends on the one-loop correction to the K\"ahler potential.
At the minimum/maximum of the potential, the $b$ mass-squared is given by
\beqa
m_b^2&=& \pm 3^5 e^{-2k\pi R} {\vert W_0\vert^2} M_4^2\,
\frac{\Omega^\prime +\Omega^{\prime\prime}}{ (\Omega\Omega^{\prime\prime}
-{\Omega^\prime}^2)^2}\nonumber\\ 
&=& \pm 3 e^{2k\pi R} \frac{\vert W_0\vert^2}{M_4^6} \, (1-e^{-2\pi kR})^4
\, (\Delta\Omega^\prime +\Delta\Omega^{\prime\prime})
\,,
\eeqa
where we again used the fact that 
$\Omega^\prime +\Omega^{\prime\prime}=0$ at tree level
and substituted $\Omega=\Omega_{tree}$ everywhere else.

With pure gravity in the bulk, the one-loop correction 
to the K\"ahler function is given by the following 
integral~\cite{Falkowski:2005fm},
\begin{eqnarray}\label{omegagrav}
\Delta \Omega_{gravity}&=&\frac {k^2 a_\pi^2} {4 \pi^2}\,
\int_0^\infty dy~ y \log \left(1-\frac {I_1(y a_\pi) K_1(y)}{K_1(y
a_\pi) I_1(y)}\right)\nonumber \ ,
\\  
a_\pi^2&=&e^{-k \pi(\rad+\bar{\rad})}\ . \label{omegagr}
\end{eqnarray}
The $b$-dependent potential can then be written
as an integral  over a combination of modified Bessel
functions, and evaluated numerically for any value
of $kR$.
Alternatively, since we are interested in small warping, with
\beq
\frac{1}{L}\ll k \ll \frac{1}{R} \ ,
\eeq
we can expand the integrand for $exp(-2\pi kR)\sim1$.
The leading order term in this limit 
is~\cite{Rattazzi:2003rj,Falkowski:2005fm}\footnote{The next term
in the expansion is $\#\frac{k}{\rad+\bar\rad}$, where the 
coefficient can be calculated using~(\ref{omegagrav}).}
\beq\label{loopkflat}
\Delta\Omega =  -\frac{\zeta(3)}{4\pi^4} \, 
{1\over (\rad+\bar{\rad})^2}
\ .
\eeq
Using the superpotential~(\ref{super}) we then find,
\beq\label{vflat4d}
\Delta V = -
\frac{3\zeta(3)}{8\pi^6}\,\frac{1}{(k L)^2} \frac{R^2}{r^6}\,
\sin^2\left(\frac{3\pi k b}{2}-{\phi\over2}\right)\ ,
\eeq
which matches~(\ref{kkcorr}) for $b=0$ and small $\phi$.
As we mentioned at the end of the previous section, 
the one-loop, supersymmetry breaking contribution to
the potential appears at zeroth order in $kR$.
In contrast, the tree-level potential near $r=R$
is order $(kR)^2$, because it should vanish in the flat space limit.
Near the extrema with $r=R$ we then have
\beq
{\Delta V\over V_{tree}} = \frac{\zeta(3)}{8\pi^6}\,
{1\over (k R)^2}\, \frac{1}{(M_4R)^2}\, \sin^2{3\pi kb-\phi\over2}\ ,
\eeq
so that the loop suppression is partly compensated
by inverse powers of $k R$.
Thus, for small values of $kR$, the quantum correction significantly modifies
the radion potential and cosmological constant.

Let us therefore examine the full potential. For small $k$,
the tree plus loop potential is given by,
\beq
V=\frac{3M_4}{L^2}\left[\frac{1-2x}{x^2}-\frac{\zeta(3)}{8\pi^6}
\frac{1}{(M_4R)^2} \frac{1}{(kR)^2} \frac{1}{x^6} \sin^2\frac{3\pi
  kb}{2} \right]~,
\eeq
where
\beq
x=\frac rR~.
\eeq
As we saw above, the one-loop correction with pure gravity is always
negative. 
Supersymmetry is unbroken for $b=0$. This vacuum is a stable saddle 
point with a positive mass squared for the radion, and a negative mass squared 
for $b$, which is still above the Breitenlohner-Freedman 
bound~\cite{Breitenlohner:1982bm},
\beq
m_b^2 
= - {9\zeta(3)\over 8 \pi^4}\, {1\over (M_4 R)^2}\, {1\over L^2}
>-\frac{9}{4L^2} \ .
\eeq
The  4d cosmological constant and
radion VEV are not modified in this vacuum\footnote{Note that
we implicitly imposed this matching condition in eqn.~(\ref{loopkflat})
in order to determine the non-calculable contributions to
$\Delta\Omega$.}. 

For $b=1/(3k)$, supersymmetry is maximally broken.
Surprisingly, when the loop correction is sufficiently small,
the potential has {\sl two} extrema  
in the region $r<R$. One is a minimum, 
while the other is a saddle point, which is stable
for a small range of $kR$.
In either supersymmetry-breaking vacuum, the effect of the
quantum corrections is to lower the 4d cosmological constant.

We can get a positive contribution 
to the potential for a sufficient number of bulk 
hypermultiplets\footnote{The contribution of vector supermultiplets
is negative~\cite{us}.}. 
These correct the K\"ahler function by
\beq
\Delta \Omega_{hyper}=N_H \frac {k^2 a_\pi^2} {8 \pi^2}\int_0^\infty dy\, y
\log 
\left[1-\frac {I_{|c+1/2|}(y a_\pi) K_{|c+1/2|}(y)}
{K_{|c+1/2|}(y a_\pi) I_{|c+1/2|}(y)}\right]\ ,
\eeq
where $c$ is related to the bulk mass of the hypers,
and $N_H$ denotes the number of hypermultiplets.
For the special case $c=1/2$, this contribution coincides 
with eqn~(\ref{omegagrav}), up to a factor of $-1/2$.
Thus, with $N_H>2$ such hypermultiplets, the supersymmetry
breaking contribution is positive, and for small warping,
\beq\label{vhyp}
\Delta V = \left(\frac{N_H}2-1\right)\,
\frac{3\zeta(3)}{8\pi^6}\,\frac{1}{(k L)^2} \frac{R^2}{r^6}\,
\sin^2\left(\frac{3\pi k b}{2}-{\phi\over2}\right)\ .
\eeq
For a few hypermultiplets\footnote{Note that for two hypermultiplets
  with the special mass $c=1/2$, the potential remains flat to leading
  order in our expansion.}, maximal supersymmetry breaking is
always a stable saddle point of the potential,
since, again, the stability bound~\cite{Breitenlohner:1982bm} is satisfied
\beq\label{stability}
m_b^2 
> -\frac{9}{4L_{\mathrm{eff}}^2} \ .
\eeq
where $L_{\mathrm{eff}}$ is the net 4d curvature at the new minimum of the
potential. 
We then have a very interesting situation. At tree level,
the radius is stabilized due to the small detuning of the two brane tensions.
For maximal supersymmetry breaking, the negative tree-level potential
can be reduced with the radion and $b$ both stable. 
In principle, with a suitable choice of $kR$, the net 4d cosmological 
constant can be made arbitrarily small.
As long as this net cosmological constant is negative, the new vacuum is 
stable, because eqn.~(\ref{stability}) is satisfied.
For phenomenological purposes however, there is no need to set the
net cosmological constant strictly to zero. Once the MSSM is embedded
into the model, it would contribute a cosmological constant
of order $\tilde{m}^4$, where $\tilde{m}$ is the typical MSSM
soft mass. This mass would be proportional to the supersymmetry-breaking scale
$L^{-1}$. 
So the relevant question is whether the sum of the classical and quantum
contributions to the potential that we considered can be made
smaller or comparable to the MSSM contribution. 
While detailed model building is beyond the scope of this paper,
it seems that in the minimal framework we are considering, with just
$c=1/2$ hypermultiplets, this cannot be achieved.
This is related to the fact that both the supersymmetry-breaking
scale and the cosmological constant are proportional to a single scale,
$L^{-1}$, so that the form of the potential is determined by
a single dimensionless parameter, namely, $kR$.

We close this section with a comment regarding the periodicity
of the potential.
So far we have written the  potential
in terms of $b$, which has mass dimension $-1$.
Switching to the canonically normalized field,
\beq
b_c 
  = \frac{\pi{k}e^{-k\pi R}}{1-e^{-2k\pi R}}\, M_4\, b\ ,
 \eeq
the period  is roughly
$$
 \frac{{\pi}e^{-k\pi R}}{1-e^{-2k\pi R}}\, M_4\ ,
$$
which is always above the KK scale. 
In particular, in the limit $kR\ll 1$, the period of $b_c$
is $M_4/(3 kR)$, which  is much larger than the Planck scale.  
Thus, we can only use the 4d effective theory 
to study small fluctuations of $b$ around
either the supersymmetric or supersymmetry-breaking extremum.

\section{Conclusions and Outlook}
In this paper we studied a supersymmetric orbifold model with detuned 
brane tensions.
We examined the KK gravitino spectrum,
and calculated the shifts of the KK masses
due to detuning and supersymmetry breaking.
We then calculated the one-loop correction
to the potential, using both the KK spectrum and the
low-energy radion theory.

Throughout we take the detuning to be small, so
that the 4d curvature is the lowest scale in the problem.
In addition, we concentrate on the small warping case,
since in this case the loop suppression of the Casimir
energy is partly compensated by powers of $kR$.
We note that higher loops will be suppressed by the usual
loop factors compared to the one-loop contribution.
The parameter $kR$ only appears in the ratio
of the one-loop correction to the tree-level potential.
As we explained, the reason is that the loop correction
remains finite as we decrease the warping, while the tree-level
potential is proportional to the 5d curvature.

We found that both the radion and graviphoton 
are stabilized at one loop, with supersymmetry either
unbroken or maximally broken.
In the presence of a few bulk hypermultiplets,
supersymmetry breaking gives a significant positive 
contribution to the potential. 
In the minimal framework we considered, the form of the potential
is determined by the parameter $kR$.
Embedding the MSSM into this framework, the typical soft mass
would be of order $L^{-1}$ at most.
Then, even though the classical and loop contribution
can almost cancel, the net 4d cosmological constant
is never smaller than the typical cosmological constant
generated by the MSSM. 
It would be interesting to study models with hypermultiplets
of different bulk masses, where the potential is more
complicated and there is more freedom in the choice of parameters.

While we have only studied here the two-brane detuned system,
it would be interesting to explore the supersymmetric
single-brane model, with detuned brane tension.
This model is a supersymmetric generalization of the Randall-Karch
model, in which 4d gravity emerges through an ultra-light
graviton KK mode. 
The supersymmetric version of this theory therefore
looks like 4d massive supergravity~\cite{GSS} in AdS$_4$.
While the mass of the ultra-light ``graviton'' is
below the AdS$_4$ curvature,
it would be interesting
to understand the structure of this theory
from a purely theoretical point of view.

\vskip 0.15in
{\noindent \bf Acknowledgments}
\vskip 0.125in
\noindent
We thank Shinji Hirano, Lisa Randall, and Matt Schwartz
for discussions. We are especially grateful to Michele Redi for
collaboration at earlier stages of this work.
Research supported by the United States-Israel Science Foundation
(BSF) under grant 2002020.
The research of Y. Shadmi and A. Katz is also supported
by the Israel Science Foundation (ISF) under grant 29/03.
 The research of Y.~Shirman is supported by the US Department of
Energy under contract W-7405-ENG-36.
Y.~Shadmi and Y.~Shirman thank the Aspen Center for Physics
where part of this was completed. A.~Katz also thanks LANL
for hospitality.

\appendix

\section{The framework}\label{setup}
In this appendix we present the elements of the theory,
summarizing the results
of~\cite{Bagger:2002rw,Bagger:2003vc,brwilson,Bagger:2003dy}
for completeness.
The bulk action is the supersymmetric
AdS$_5$ action,
\beqa \nonumber
S_{bulk}= M_5^3 \int d^5x\sqrt G \Bigg[-\frac12 R+6k^2 +\frac
i2\tilde\Psi^i_M\Gamma^{MNK}D_N\Psi_{Ki}
\ \ \ \ \ \ \ \ \ \ \ \ \ \ \ \ \ \ \ \ \ \ \ \\
-\frac32 k\vec q\cdot\vec\sigma_i^j\tilde\Psi^i_M\Sigma^{MN}\Psi_{Nj}
-\frac14 F_{MN}F^{MN} -i\frac{\sqrt6}{16}F_{MN}(2\tilde \Psi^{Mi}
\tilde \Psi_i^N+\tilde \Psi^i_P\Gamma^{MNPQ}\Psi_{Qi})\nonumber\\ 
-\frac{1}{6\sqrt6}\epsilon^{MNPQK}F_{MN}F_{PQ}B_K +\frac{\sqrt6}{4}
k \vec q\cdot \vec\sigma^j_iB_N\tilde
\Psi^i_M\Gamma^{MNK}\Psi_{Kj}\Bigg]~,\ \ \ \ \ \ \ \ \ \nonumber
\eeqa
where $M_5$ is the 5d Planck scale, and $k$ is the 5d bulk 
cosmological constant.
The unit vector $\vec q$ defines the  gauged $U(1)$ subgroup of 
$SU(2)_R$ and we will set it to $(0,0,1)$ for simplicity.
The graviphoton, $B_M$, is the gauge boson of this $U(1)_R$.
Because of the 5d curvature, the gravitino is charged under the $U(1)_R$.
Note that a non-zero $B_5$ gives a non-zero mass to the 4d gravitino.
We also define for convenience:
\beq
T=6 M_5^3 k.
\eeq
Apart from brane tensions, 
the brane action contains gravitino mass terms, 
\beqa
S_{brane}= &-&\int d^4x dx_5\, e_4 \,[T_0
+ 2\alpha_0 (\psi_{m1} \sigma^{mn} \psi_{n1} + h.c.)] \, \delta(x_5)\\
\nonumber
&-&\int d^4x\, dx_5\, e_4 \, [T_\pi
- 2\alpha_\pi (\psi_{m1} \sigma^{mn} \psi_{n1} + h.c.)]\,  \delta(x_5-R)\ .
\eeqa
In order for the bulk plus branes action to be locally supersymmetric,
the gravitino  brane mass terms must satisfy
\beq\label{branemass}
T_i= \frac{
  1- |\alpha_i|^2 }{1 +|\alpha_i|^2} T~.
\eeq
Therefore, the absolute values of the brane tensions cannot
be bigger than the bulk cosmological constant:
\beq
|T_{0, \pi}| \leq T~.
\eeq
The inequality is saturated for the tuned RS case.

The resulting 4d theory can be either AdS$_4$ or Mink$_4$,
with the metric,
\beq\label{metrica}
ds^2=a^2(x_5)\hat g_{\mu \nu}dx^\mu dx^\nu-dx_5^2 \ ,
\eeq
where $\hat g_{\mu \nu}$ denotes the standard AdS$_4$ or Mink$_4$ metric
in Poincare coordinates, and
where the warp factor is given by
\beq
a(x_5)=e^{-k x_5}+\frac{1}{4 k^2 L^2}e^{k x_5}~.
\eeq
Here $L$ is the 4d curvature radius, given by
\beq\label{4dcurv}
\frac{1}{4 k^2 L^2}=\frac{T-T_0}{T+T_0}~.
\eeq
The brane distance is
\beq\label{radius}
R=\frac{1}{2k\pi}
\ln  \frac{(T+T_0)(T+T_\pi )}{(T-T_0)(T-T_\pi )}~,
\eeq
and the 4d Planck scale is
\beq
M_4^2= {M_5^3\over k} (1- e^{-2k\pi R})\ .
\eeq
Note that the requirement of local supersymmetry
excludes the dS$_4$ case, since the latter implies
$T_0 >T$, in conflict with~(\ref{branemass}).

The condition~(\ref{branemass}) restricts just the magnitudes
of the the gravitino brane mass terms, but not their phases.
When $\alpha_0$,  $\alpha_\pi$  have different phases,
there is no solution to the Killing spinor equations (valid in the bulk
{\it and} on the branes) and supersymmetry is spontaneously
broken.
Actually, this is only true for $B_5=0$.
Allowing some constant $B_5$ background, the condition for
unbroken supersymmetry becomes,
\beq\label{b5}
\alpha_\pi = \alpha_0 e^{k \pi ( R - i\sqrt6  B_5)} \ . 
\eeq
Writing $\alpha_i= \vert\alpha_i\vert e^{i\phi_i}$,
we see that the phase difference
$$\phi=\phi_0-\phi_\pi\ ,$$
can be compensated by a shift of $B_5$, as expected from
the $U(1)_R$ invariance of the 5d theory.

For the tuned case, $T_0=-T_\pi=T$, we recover RS1, with zero
4d curvature, and with the radius $R$ undetermined.
Furthermore, the gravitino mass terms $\alpha_0$ and $\alpha_\pi$
 vanish in this case,
so that equation~(\ref{b5}) is always satisfied,
and supersymmetry is preserved.

To conclude this appendix we relate the brane distance $R$ 
and $B_5$ to the radion superfield $\rad=r+i b$ used in 
section~\ref{4dsection}:
the graviphoton Wilson line is defined as
\beq
\label{graviphoton}
b = {1\over\sqrt6\pi} \int_{-\pi R}^{\pi R} B_5\, dx_5\ ,
\eeq
and the radion
\beq
r={1\over2\pi R}\, \int_{-\pi R}^{\pi R}\, \sqrt{G_{55}}\, dx_5 \ .
\eeq

\section{The KK mass shift}\label{shifts}
In this appendix we derive the mass shifts of the gravitini KK modes by
expanding around the tuned RS case. We will do so by solving the
appropriate Schr\"odinger problem for the gravitini.
To write down the gravitini equations of motion,
we rescale the 4d coordinates as\footnote{With this rescaling,
we reproduce the conventions of~\cite{Bagger:2003vc}.}
\beq
\tilde x^\mu=
\frac{2T}{T+T_0}\,
x^\mu\,.
\eeq
At the end of the day, we will therefore need to rescale the masses
we find by
\beq\label{rescm}
m \to \frac{2T}{T+T_0}\, m\ .
\eeq
The gravitini equations of motion are then
\beqa\label{eom1}
\tilde a\frac{db_1}{dr} +\frac{3k}{2}\tilde ab_1 +\frac{d\tilde
  a}{dr}b_1&  =&m\bar
b_2\\\label{eom2} 
\tilde a\frac{db_2}{dr}-\frac{3k}{2}\tilde ab_2+\frac{d\tilde
  a}{dr}b_2& =&-m\bar b_1~,
\eeqa
where $b_1$, $b_2$ are gravitini wave-functions and
\beq
\tilde a(r)= \frac{T+T_0}{2T}\,
 a(r)
\,,
\eeq
with the boundary conditions:
\beqa\label{bc1}
b_2(r=0)&=&\alpha_0 b_1(r=0)\\\label{bc2}
b_2(r=\pi R)&=&\alpha_{\pi} b_2(r=\pi R)~.
\eeqa
Note that $R$ in these equations is the radius given by eqn.~(\ref{radius}).

To convert these equations to a Schr\"odinger-like problem we go to
coordinates:
\beq
z(r)= \frac{2}{kA}\arctan \left(\sqrt{\frac{T-T_0}{T+T_0}}
e^{kr}\right)~,
\eeq
where
\beq
A^2=1-\frac{T_0^2}{T^2}\ ,
\eeq
and rescale 
\beq
b_i(z)=\frac{1}{kA}\sin (kAz)\psi_i (z) \ .
\eeq
The $\psi_2$ equation of motion then takes the form
\beq
\Bigl(-\partial_z^2 +V_2(z)\Bigr)\psi_2(z) =m^2 \psi_2(z)~,
\eeq
with
\beq\label{fullpotential}
V_2(z)=-\frac{3k^2A^2}{2}\frac{\cos(kAz)}{\sin^2(kAz)}
+\frac{9k^2A^2}{4}\frac{1}{\sin^2(kAz)}~.
\eeq  
The second wavefunction is completely determined by  
equation~(\ref{eom1}) as
\beq\label{match}
\bar\psi_1=-\frac{\dot
  \psi_2}{m}+\frac52\frac{\psi_2}{mz}+\frac{\psi_2}{m} \frac{d\ln
  a}{dz}~.
\eeq
For completeness we also write down the profile function in these 
new coordinates:
\beq
a(z)=\frac{A}{\sin(kAz)}~.
\eeq

Since we want to perturb around the tuned case, it is
useful at this stage to define the following quantities
\beqa
\epsilon&\equiv& \sqrt{\frac{T-T_0}{T}}\\
\delta&\equiv& \sqrt{\frac{T-T_\pi}{T}}\,,
\eeqa
and we require both $\epsilon\ll 1$ and
$\delta\ll 1$. Note
also that 
\beq
\delta =e^{\pi kR}\epsilon\,,
\eeq 
and we keep $R$ finite.

Expanding the potential $V_2(z)$ in powers of $\epsilon$,
we solve the equations of motion for $\psi_2$. 
Writing $\psi_2$ as an
expansion in $\epsilon$, we get  
\beq
\psi_2=G(z^{1/2}J_1(mz)+\epsilon^2
\frac{k^2z^{3/2}}{m}J_2(mz)-\epsilon^2
\frac{k^2}{m^2}z^{1/2}J_1(mz) +H(J\leftrightarrow Y)~,
\eeq
where $G$ and $H$ are constants, which must be determined 
from the  boundary
conditions. 
Using the matching condition~(\ref{match}) we also have
\beq 
\psi_1=\bar G(\sqrt zJ_2(mz)+\epsilon^2 \frac{k^2}{m^2}\sqrt
zJ_2(mz)-\frac{\epsilon^2}{3} \frac{k^2z^{3/2}}{m}J_1(mz))+\bar H
(J\leftrightarrow Y)~.
\eeq
Expanding also $R$, $\alpha_0$ and $\alpha_\pi$ we get a system of
four linear 
homogeneous equations for $G$ and $H$ . The masses can be computed
from the requirement that 
the characteristic determinant of the system should
vanish.   

Finally we should rescale the results by~(\ref{rescm}). 
Writing
\beq
\frac{{m^{(n)}}^{\pm}}{k}= \frac{m^{(n)}_0}{k} \pm \frac{1}{kL} c^{(n)}_1  
\pm \frac{1}{kL} c^{(n)}_{1,SB} \phi^2
+ \frac{1}{(kL)^2} c^{(n)}_2  + \frac{1}{(kL)^2}c^{(n)}_{2,SB} \phi^2~, 
\eeq
we find that the shift due to detuning, with no 
supersymmetry breaking, is $n$-independent, 
\beq\label{startc}
c^{(n)}_1= \frac12 \ ,
\eeq
while the leading order shift due to supersymmetry breaking is
\beq
c^{(n)}_{1,SB}= -\frac14  
\frac{e^{k\pi R}\Delta_{12}^{(n)}\Delta_{21}^{(n)}}
    {(e^{k \pi R}\Delta_{12}^{(n)}+\Delta_{21}^{(n)})^2}\,.
\eeq
where
\beq
\Delta_{ij}^{(n)}\equiv\left|\begin{array}{cc}
J_i({c^{(n)}_0}) & Y_i({c^{(n)}_0})\\
J_j(e^{k \pi R}{c^{(n)}_0} ) & Y_j( e^{k\pi R} {c^{(n)}_0 })
\end{array}\right|~,
\eeq
where $c^{(n)}_0=m^{(n)}_0/k$.
Note that
$\Delta_{11}^{(n)}=0$.
At the next order
\beq\label{seconddeviation}
c_2^{(n)} = -{1\over {12\sqrt2}}\,
\left[
{e^{3k\pi R} \Delta^{(n)}_{12}  
+4 \Delta^{(n)}_{21} +3 e^{k\pi R}   
\over
  e^{k\pi R} \Delta^{(n)}_{12} +\Delta^{(n)}_{21}}\, c_0^{(n)}-3\sqrt2
c_0^{(n)} 
-{3\over c_0^{(n)}}\right]\ , 
\eeq
and
\beq\label{endc}
c_{2,SB}^{(n)} = - {e^{k\pi R} \Delta^{(n)}_{12}\Delta^{(n)}_{21}
        (-3 e^{k\pi R} \Delta^{(n)}_{12} -3\Delta^{(n)}_{21}
         +2 e^{k\pi R} c_0^{(n)}\Delta^{(n)}_{22}) \over
4\sqrt2(e^{k\pi R}
\Delta^{(n)}_{12}+\Delta^{(n)}_{21})^3}\, {1\over c_0^{(n)}}~. 
\eeq

At the lowest level, one gravitino state is projected out
by the orbifolding, and we are left with a gravitino
with two degrees of freedom, and a  radion fermion with two degrees
of freedom, both of mass
\beq
m=\frac1L  .
\eeq
When supersymmetry is broken, the radion fermion is eaten and
the 4d gravitino has four degrees of freedom with
mass
\beq
\label{zeroshift}
m_{zm}= \frac1L+\frac{1}{2L}\, \frac{e^{2\pi kR}}{(e^{2\pi
    kR}-1)^2} \, \phi^2~.
\eeq
Using (\ref{zeroshift}) we can define 
\beq
c_1^{(0)}=1, ~~~~c_{1,SB}^{(0)}= 
\frac{e^{2\pi k R}}{2(e^{2\pi k R}-1)^2}
\eeq
Since the mass of this mode is zero in the tuned case, its contribution
to the vacuum energy starts at order $1/L^2$,
so we don't need higher corrections.
   
\section{Casimir energy in flat limit}\label{flat}
Here we evaluate the correction~(\ref{vacen}) for small warping.
In this limit,
\beq
c_0^{(n)}=\frac{n\pi }{e^{\pi kR}-1}+\frac{3(e^{\pi kR}-1)}{8n\pi
  e^{\pi kR}}+{\cal O}((e^{\pi kR}-1)^3)~,
\eeq
where the leading term is just the flat space result,
and
\beqa
  c_1^{(n)}c_{1,SB}^{(n)}&\to&-\frac{ e^{\pi
      kR}\phi^2}{4(e^{\pi kR}-1)^2}+{\cal O}(1)\\
  c_0^{(n)}c^{(n)}_{2,SB}&\to& {\cal O}(1)~.
\eeqa
Thus the leading order of~(\ref{vacen}) gives the dominant
contribution.
Substituting these results in~(\ref{vacen}) we have:
\beqa
\Delta V=-\frac{1}{(kL)^2}\frac{2}{(\pi
  R)^2}\,\phi^2\,\cdot\ \ \ \ \ \ \ \ \ \ \ \ \ \ \ \ \ \\
\int\frac{d^4p}{(2\pi)^4}\Biggl[
\frac12 \sum_{n=1}^{\infty}\frac{1}{(p^2+(n/R)^2)}
&-&\sum_{n=1}^{\infty}\frac{(n/R)^2}{(p^2+(n/R)^2)^2}
+ \frac{1}{4p^2}\Biggr]~.\nonumber
\eeqa
The last term is the contribution of the zero mode, and
cancels the quadratic divergence from the first two
sums. Performing the sums by standard techniques\footnote{See for
  example~\cite{Delgado:1998qr}.} we obtain:
\beq\label{kkcorrabs}
\Delta V = - {3 \zeta(3)\over
2^5\pi^2}
{1\over (\pi R)^4}\, {1\over (k L)^2}\, \phi^2\ .
\eeq                              


\end{document}